\documentclass[sigplan,10pt,nonacm]{acmart}
\setcopyright{none}
\acmDOI{}
\acmISBN{}
\acmPrice{}

\usepackage{algorithmic}
\usepackage{graphicx}
\usepackage{textcomp}
\usepackage{xcolor}
\definecolor{revisionpurple}{RGB}{128,0,160}
\definecolor{revisiongreen}{RGB}{0,128,64}
\usepackage{booktabs}
\usepackage{multirow}
\usepackage{subcaption}
\usepackage{placeins}
\usepackage{tikz}
\usetikzlibrary{arrows.meta, positioning, shapes.geometric, decorations.pathreplacing}

\def\BibTeX{{\rm B\kern-.05em{\sc i\kern-.025em b}\kern-.08em
    T\kern-.1667em\lower.7ex\hbox{E}\kern-.125emX}}

\newcommand{\FFTTwo}{\mathrm{FFT2}}

\begin{document}
\title{Spectral Analysis for Sparse Matrix Computation: Insights and Potential}

\author{Ruifeng Zhang}
\affiliation{%
  \institution{North Carolina State University}
  \city{Raleigh}
  \state{North Carolina}
  \country{USA}
}

\author{Xipeng Shen}
\affiliation{%
  \institution{North Carolina State University}
  \city{Raleigh}
  \state{North Carolina}
  \country{USA}
}

\begin{abstract}
	Sparse computations are fundamental to scientific computing, graph analytics, and machine learning, yet their performance is highly sensitive to the diverse sparsity and patterns. 
    This is because cache reuse, memory coalescing, and load balancing depend critically on the sparsity patterns. This work gives the first known exploration of the connections between sparse matrix computation and spectral analysis by treating sparse matrices as two-dimensional signals and analyzing their frequency-domain representations through Fast Fourier Transform. We show that spectral signatures uncover global structural characteristics that are not sufficiently captured by conventional spatial statistics and provide complementary information for understanding sparse computation performance. 
	    Experiments on incorporating spectral features into machine-learning-based SpMV format selection demonstrate the usefulness of such spectral analysis over a state-of-the-art spatial-only model. By uncovering the principled connections between spectral characteristics and sparse matrix computations, this work introduces a novel analytical perspective into sparse computation, and provides a new approach to enhancing the current sparse structure characterization and optimization. On pruned LLM decoding, adding spectral features improves kernel selection and yields 1.035--1.245$\times$ kernel speedups.
\end{abstract}

\ccsdesc[500]{Computer systems organization~Parallel architectures}
\ccsdesc[300]{Computing methodologies~Machine learning}

\keywords{sparse matrices, sparse kernel selection, Fast Fourier Transform,
spectral analysis, GPU performance}

\maketitle


\section{Introduction}
\color{black}Sparse computation is essential for various applications, such as scientific simulations~\cite{davis2010algorithm,lu2020efficient}, Graph Neural Networks (GNNs)~\cite{wang2019deep,huang2020ge}, and Deep Neural Networks (DNNs) and modern LLM-based AI~\cite{molchanov2017variational,zheng2022sparta,roy2021efficient, dao2022flashattention}. 
Unlike General Matrix-Matrix Multiplication (GeMM), which exhibits regular access patterns, sparse computations are challenging for the highly diverse nonzero distributions in sparse matrices. Such irregularity often leads to irregular memory access, poor memory coalescing, and workload imbalance on parallel architectures~\cite{gale2020sputnik}. As a result, characterization of sparse matrices and understanding of sparse patterns are crucial for maximizing the performance of sparse computations. For instance, numerous studies have shown that the efficiency of sparse computations, especially on massively parallel architecture (e.g., GPUs), depends heavily on the selection of the right storage format and kernel implementation, which are critically determined by the matrix's sparse patterns~\cite{wen2016learning,zhou2021learningnm,xia2023sheared}.

\color{black}
Existing works~\cite{zhou2019enabling,dufrechou2021selecting,yesil2023wise} have concentrated on spatial characterizations of sparse matrices, such as dimensions, sparsity, and row/column nonzero distributions. Spatial features effectively capture certain pattern information, but they often miss global structures. Matrices with similar row-length distributions can still arrange their nonzeros into different bands, blocks, or mixed structures, changing locality, memory coalescing, and load balance. Consequently, distributional summaries alone may not provide enough information to select the best storage format and kernel implementation.

This work gives the first known study on a different way of sparse matrix characterization, spectral analysis. Spectral analysis is an essential technique in signal processing and image understanding. A common approach is to apply Fast Fourier Transform (FFT) to the signal to produce the magnitude and phase maps of the various frequency components in the signal. It is useful in revealing the global periodicity, orientation, and scale of the data in the signals~\cite{cooley1965fft,oppenheim1999discrete,uzun2005fpga,zhang2022fast}. The key insight driving our exploration is that if we view a sparse matrix as a binary 2D image, where zero elements becomes black pixels and nonzero elements pixels with value equaling 1, we may be able to obtain spectral features of sparse matrices through the spectral analysis in the signal processing field. In signal processing, utilizing frequency-domain representations are often more expressive than pure spatial statistics~\cite{chen2021few,balcilar2020bridging}. We are curious of whether applying spectral analysis to sparse matrices can provide additional insights into the relationship between structural patterns of sparse matrices and the performance of their computations. To our best knowledge, spectral analysis of sparse matrices have not been systematically explored before. The closest efforts are the application of spectral analysis to graphs, which applies Graph FFT or spectral filtering defined in the eigenbasis
of the graph Laplacian to graphs~\cite{von2007tutorial,spielman2012spectral,shuman2013emerging,bruna2013spectral,kipf2016semi,zhao2019t,he2021community}. In contrast, our approach applies a 2-D FFT directly to
the nonzero pattern of a sparse matrix by treating it as an image. Even though some sparse matrices originate from graphs (e.g., adjacency matrices that are symmetric), our approach applies to general sparse matrices. 

Our exploration focuses on the following open research questions:


\noindent
\textbf{RQ1: How do spectrum signatures reflect sparsity patterns?}
Spectral features are most useful if they can reflect specific sparse structures. 
Section~\ref{sec:interpretation} uncovers the connection through spectral analysis of a set of pure and composite representative sparse patterns through matrix synthesis. This establishes an intuitive and empirical basis for interpreting frequency-domain characteristics and their connections with spatial sparsity structures.

\noindent
\textbf{RQ2: How do the connections manifest in real-world sparse matrices? Can they help sparse matrices analysis?}
Section~\ref{sec:dataset-features} examines the spectral features of 1,314 real-world sparse matrices from the SuiteSparse collection, confirming that the connections between sparse patterns and spectral signals are stable and consistent across real-world matrices. It further gives the first demonstration of the effectiveness of spectral features for sparse matrix analysis by applying them to matrix clustering.

\noindent
\textbf{RQ3: Can spectral features help improve sparse computation performance?}
Section~\ref{sec:application} answers the question by applying spectral features to SpMV storage-format and kernel selection, where even modest gains in prediction accuracy can translate into meaningful runtime savings at scale. Across SuiteSparse and pruned-LLM matrices, spectral features complement traditional spatial features and improve prediction accuracy and selected-kernel runtime.

\color{black}
Our main contributions are as follows:
\begin{itemize}
	\item We introduce the first 2-D FFT-based view of sparse matrices that turns nonzero patterns into interpretable spectral signatures. Through controlled synthetic examples, we show how common structures such as blocks, bands, and their combinations map to stable characteristics in the frequency domain.
	\item We propose a density map normalization technique to mitigate spectral degradation caused by intra-structure sparsity and noise, enabling robust feature extraction from real-world matrices.
	\item We demonstrate that real-world matrices exhibit stable and interpretable spectral signatures, with clustering in FFT feature space revealing meaningful structural definitions and optimization potentials.
		\item We showcase the practical value of these features for SpMV kernel selection on SuiteSparse and pruned-LLM matrices. Adding spectral features to the spatial baseline improves prediction accuracy and delivers up to 2.75$\times$ speedup over a fixed SpMV baseline.
\end{itemize}
\color{black}

\begin{figure}[htbp]
	\centering
	\includegraphics[width=0.92\linewidth]{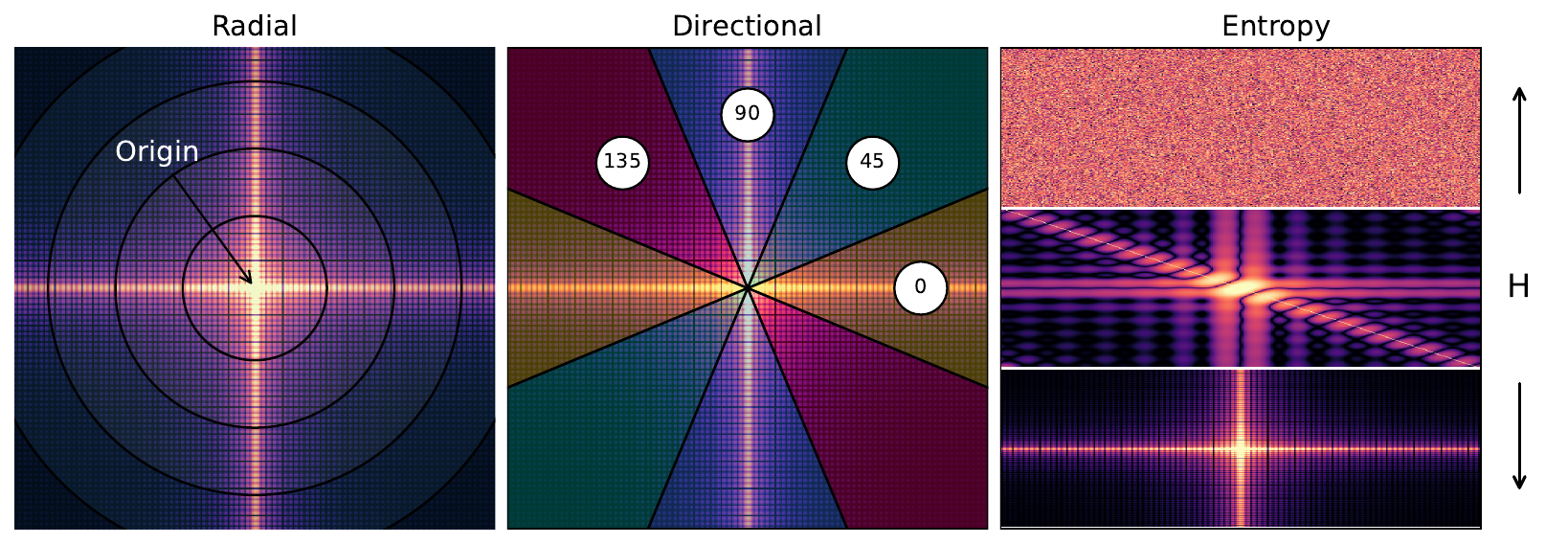}
	\Description{Schematic of radial rings, directional sectors, and entropy over FFT magnitude spectra.}
	\caption{\color{black}Illustration of the three FFT features used in this work: Radial energy, Directional energy, and spectrum Entropy. The background images for Radial and Directional, as well as the images used for Entropy, are FFT magnitude spectra obtained by shifting the zero-frequency component to the origin and applying log scaling.\color{black}}
	\label{fig:fft-feature-schematic}
\end{figure}

\section{Background}\label{sec:background}

Fast Fourier Transform (FFT) is an efficient algorithm that computes the Discrete Fourier Transform (DFT), transforming a discrete signal from the time or spatial domain into the frequency domain by decomposing it into a set of sinusoidal components with different frequencies, amplitudes, and phases. The 2-D FFT maps a spatial signal $D \in \mathbb{R}^{m \times n}$ to complex
frequency coefficients $F = \FFTTwo(D)\in \mathbb{R}^{m \times n}$, where magnitude captures the energy at
each spatial frequency and phase captures spatial alignment. Concretely, the DFT is
\begingroup
\begin{equation}
	\begin{aligned}
		F(u,v) & = \sum_{i=0}^{m-1} \sum_{j=0}^{n-1} D_{i,j}
		\, e^{-2\pi i\!\left(\frac{ui}{m} + \frac{vj}{n}\right)}, \\
		M(u,v) & = |F(u,v)|, \\
		L(u,v) & = \log(1 + M(u,v))
	\end{aligned}
\end{equation}
\endgroup
Where $M$ is the magnitude spectrum and $L$ is the log spectrum. $L$ is typically visualized as the log-magnitude spectrum of the DFT after applying an FFT shift to center the zero-frequency component. Figure~\ref{fig:canonical-fft}
illustrates how different sparse structures are translated to different log spectra.

Since the FFT produces a magnitude matrix of the same size as the original matrix, it is common to extract compact and informative spectral descriptors from the frequency domain. Let $p_{u,v}$ denote the normalized spectral mass
\[
	p_{u,v} = \frac{M(u,v)}{\sum_{u,v} M(u,v)}
\]
We can derive the following features:

\textbf{Radial energy} characterizes structural scale by grouping frequencies into concentric rings $\mathcal{R}_k$  (Figure~\ref{fig:fft-feature-schematic}, left):
\begingroup
\begin{equation}
	E_r(k) = \sum_{(u,v)\in\mathcal{R}_k} p_{u,v}
	\label{eq:radial_energy}
\end{equation}
\endgroup
Concentration in low-frequency rings (i.e., those close to the origin) indicates large-scale regularity (blocks,
	      broad bands), whereas high-frequency mass indicates fine-grained or
	      irregular structure. 
          
\textbf{Directional energy} captures orientation by partitioning the spectrum into four angular sectors $\Theta_\ell$ (Figure~\ref{fig:fft-feature-schematic}, center) :
\begingroup
\begin{equation}
	E_\theta(\ell) = \sum_{(u,v)\in\Theta_\ell} p_{u,v}.
	\label{eq:directional_energy}
\end{equation}
\endgroup
It captures oriented structures. Row-aligned blocks tend to produce
	      horizontal/vertical energy, whereas diagonal and anti-diagonal bands produce
	      $45^{\circ}$/$135^{\circ}$ ridges.
          
\textbf{Spectral entropy} quantifies the concentration versus diffusion of spectral energy (Figure~\ref{fig:fft-feature-schematic}, right):
\begingroup
\begin{equation}
	H = -\sum_{u,v} p_{u,v} \log p_{u,v}
	\label{eq:spectral_entropy}
\end{equation}
\endgroup
 Highly regular patterns yield low entropy
	      (energy concentrated in a few frequency bins), while diffuse,
	      noise-like patterns yield high entropy.


\section{RQ1: Connections between Sparse Patterns and Spectral Features}\label{sec:interpretation}

To uncover how spectral features relate to sparse patterns, we use matrix synthesis to generate matrices with clear representative sparse patterns and various levels of noises. This practice gives us a full control of the matrix patterns, helping reveal the connections with spectral features. Our observations on the effects of noises in sparse patterns lead to our proposal of {\em density-map normalization}, a way to mitigate the scale challenges of matrices and local noise influence. This section reports those observations and proposals.
It should be noted that the block and band patterns form an interpretable analytic basis, not an exhaustive taxonomy or a classifier for arbitrary matrices.



\subsection{Canonical patterns and corresponding signatures}
\begin{figure}[htbp]
	\centering
	\includegraphics[width=\linewidth]{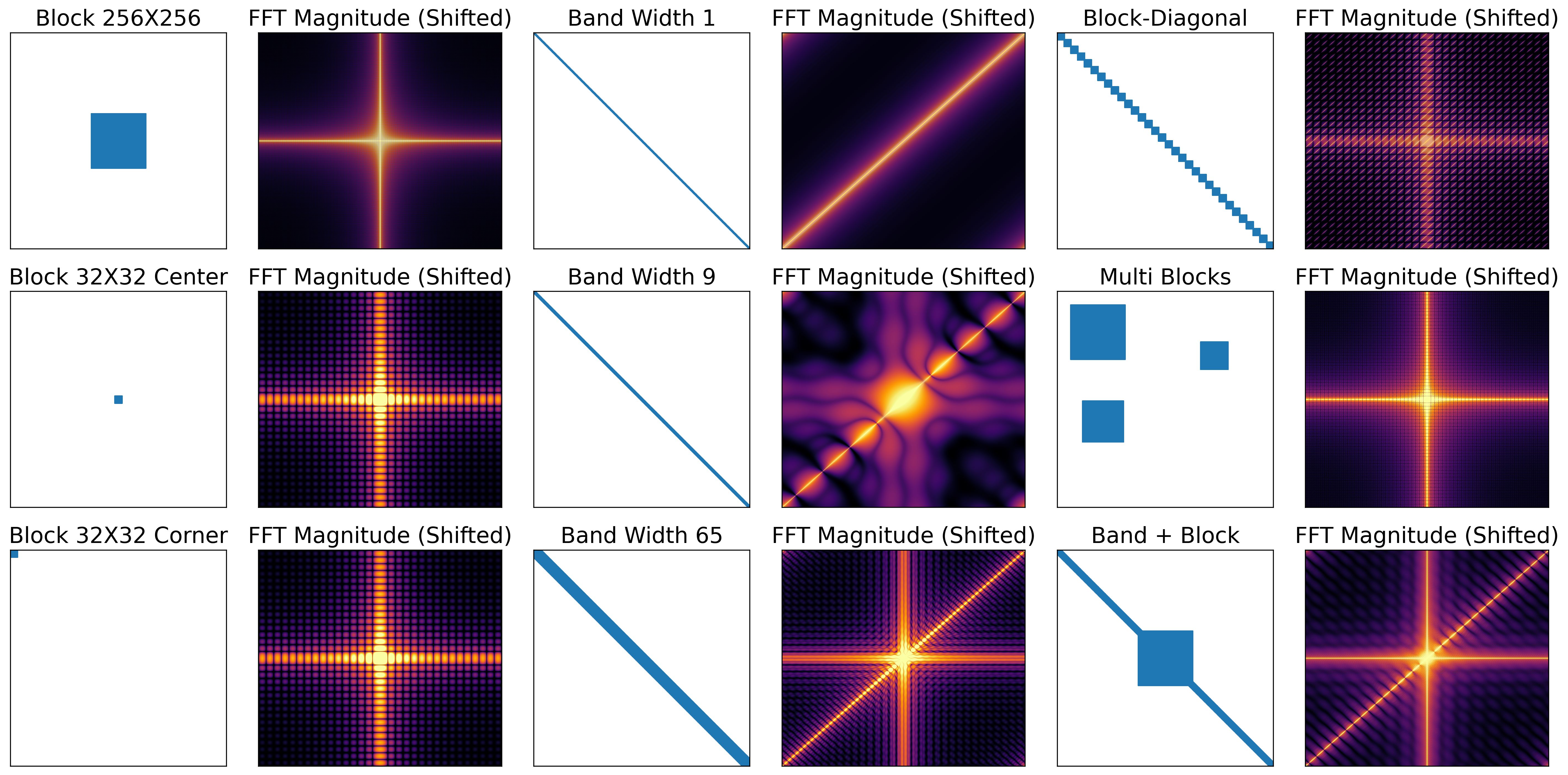}
	\Description{Six synthetic sparsity patterns paired with their shifted log-magnitude FFT spectra.}
	\caption{Canonical sparsity patterns (left) and their corresponding FFT log-magnitude spectra after shifting zero-frequency component to the center (right). The sparse matrices size is $1024\times1024$. For visualization clarity, a Gaussian smoothing is applied to the spectra to enhance thin high-energy structures without altering the underlying frequency characteristics.}
	\label{fig:canonical-fft}
\end{figure}

We start with two common sparse patterns. 

\textbf{Dense blocks.}
Figure~\ref{fig:canonical-fft} (left column) shows synthetic sparse matrices of size $1024\times1024$ with common structural patterns, including a large centered dense block, a small centered dense block, a small corner block, and diagonal bands of varying widths. A centered block concentrates spectral energy near the origin, with the energy distributed along the horizontal and vertical frequency
axes in a \emph{cross pattern}. 
Consider a sparse matrix $A$ with a centered rectangular block as
\[
	A(x,y)=f\!\left(\frac{x}{w_x}\right)
	f\!\left(\frac{y}{w_y}\right),
\]
where $f(t)$ is the rectangular function defined as
\[
	f(t)=
	\begin{cases}
		1, & |t|\le \tfrac{1}{2}, \\
		0, & \text{otherwise}.
	\end{cases}
\]
Here, $w_x$ and $w_y$ denote the block widths along the $x$- and $y$-axes, respectively.
Under the Fourier transform convention
\[
	F(u,v)=\iint A(x,y)\,e^{-2\pi i(ux+vy)}\,dx\,dy,
\]
this derives the following (details in Appendix~A):
\[
	F(u,v)=w_x w_y \,\operatorname{sinc}(w_x u)\,\operatorname{sinc}(w_y v),
\]
\[
	\operatorname{sinc}(t)=\frac{\sin(\pi t)}{\pi t}.
\]

The spectrum attains its maximum at the zero frequency $(u,v)=(0,0)$ and exhibits the cross pattern.
Increasing $w_x$ or $w_y$ narrows the main lobe in the corresponding frequency direction,
thereby concentrating more energy along the horizontal and vertical frequency axes.

The first and second rows of the left column also show that changing the block size from $256\times256$ to $32\times32$ changes the width of the central cross, while keeping the same overall low-frequency-centered signature. If the block is moved to $(x_0,y_0)$, the Fourier transform satisfies
\[
	A(x-x_0,y-y_0)
	\;\Longleftrightarrow\;
	F(u,v)\,e^{-2\pi i(ux_0+vy_0)},
\]
which introduces only a phase factor. Consequently,
\[
	\left|F(u,v)\,e^{-2\pi i(ux_0+vy_0)}\right| = |F(u,v)|,
\]
showing that the magnitude spectrum is invariant to the position of the block within the boundaries.

\textbf{Diagonal bands.}
As shown in Figure~\ref{fig:canonical-fft}(middle two columns), a thin diagonal band corresponds to an oriented ridge perpendicular to the band direction, which we term the \emph{anti-diagonal ridge}. The detailed mathematical explanation is in Appendix~B. As the band becomes wider, the spectrum exhibits a wide cross pattern alongside the original anti-diagonal ridge, with more energy centered near the zero frequency core. When the band is sufficiently thick, the spectrum reveals both anti-diagonal ridge and cross pattern, which blends band-like and block-like signatures.


\subsection{Composite Patterns}


We next synthesize matrices containing mixtures of structures to examine how their spectral signatures combine, illustrated by the right two columns of Figure~\ref{fig:canonical-fft}.  First, we construct a block-diagonal matrix composed of small dense blocks aligned along the main diagonal. Such patterns commonly arise when clustered communities coexist with local neighbor interactions. The resulting magnitude spectrum exhibits a cross-shaped energy concentration characteristic of block structures. Although each block individually produces a similar low-frequency cross pattern, placing blocks at different spatial locations introduces distinct phase shifts in the spectral domain. While phase shifts do not alter individual magnitudes, the summation of complex spectra before taking magnitudes leads to interference effects, producing faint horizontal and vertical dotted patterns and a slightly more diffused energy distribution compared to a single block.

Second, we synthesize matrices with multiple dense blocks of varying sizes and locations, representing sparse matrices with heterogeneous local structures. Despite differences in scale and position, the magnitude spectrum continues to display a prominent cross-shaped pattern, indicating that block-like structures dominate the spectral signature.

Third, we combine a diagonal band with a dense block, modeling the coexistence of large clustered regions and strong nearest-neighbor dependencies. The resulting spectrum preserves both the anti-diagonal ridge associated with the band structure and the cross-shaped low-frequency concentration characteristic of blocks, effectively blending directional and isotropic components.

Across these cases, the FFT magnitude spectrum behaves approximately as a combination of the constituent signatures, demonstrating that radial and directional features can effectively capture composite sparse patterns.

\subsection{Effects of intra-structure sparsity and noise}\label{sec:noise}
Real-world matrices rarely contain perfectly dense sub-structures; blocks or band regions typically contain some zero values, and there are usually scattered nonzeros outside the dominant structure. Both effects could diffuse the spectral energy and mask the signatures, preventing accurate detection of underlying structures. We study the effects by adding controlled noise.

\begin{figure}[htbp]
	\centering
	\includegraphics[width=\linewidth]{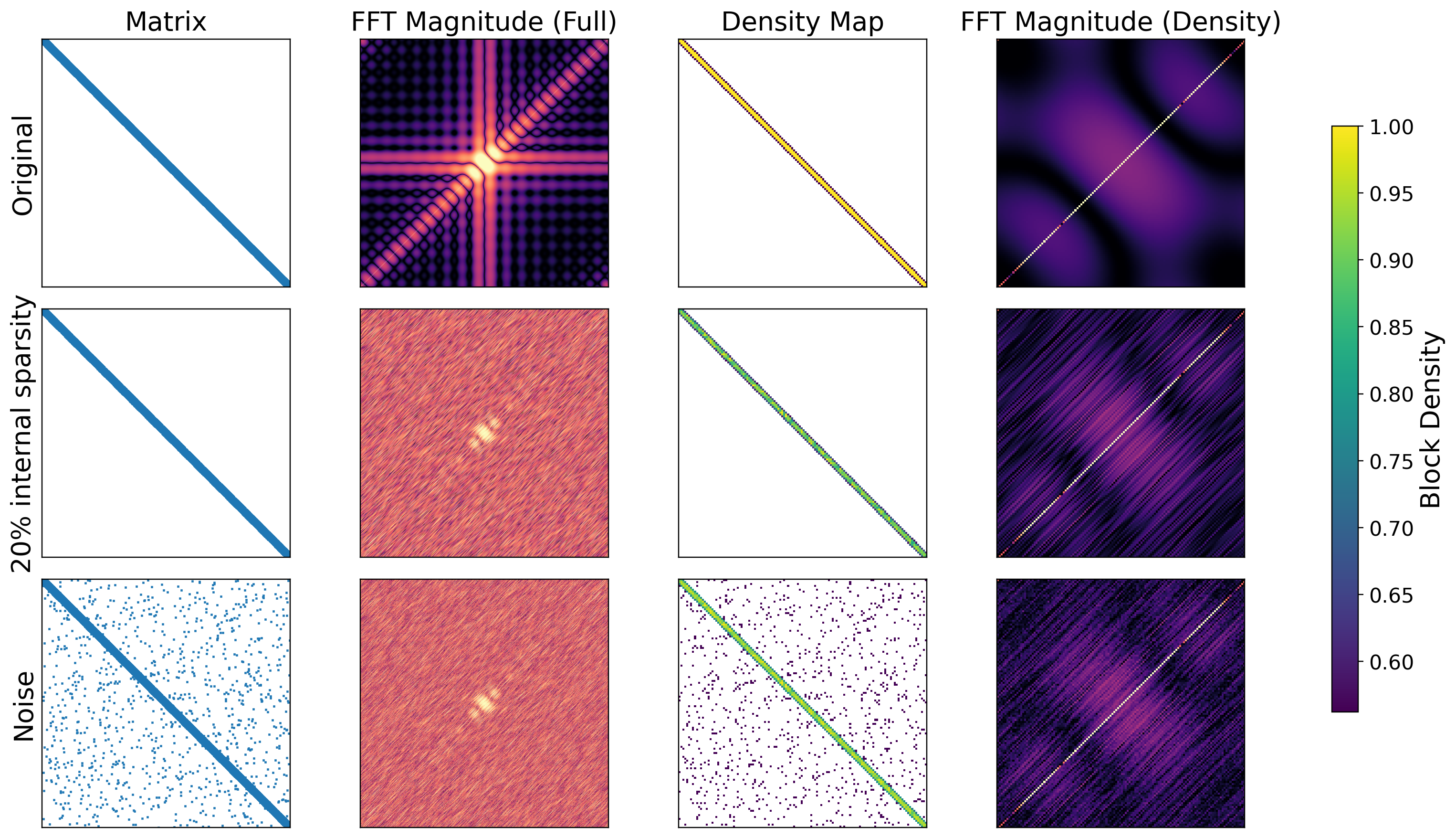}
	\Description{A diagonal-band matrix and spectrum under controlled deletion and noise, before and after density-map aggregation.}
	\caption{Effect of intra-structure sparsity and noise on a sparse matrix with diagonal band (width=33). The original matrix size is $1024\times1024$ and the density map size is $128\times128$.}
	\label{fig:sparsity-noise-fft}
\end{figure}

As shown in Figure~\ref{fig:sparsity-noise-fft}, the first two columns illustrate spectral degradation under structural sparsification and noise. Consider a diagonal band of width $33$ (top row), which exhibits a clear combination of a low-frequency cross pattern and an anti-diagonal ridge.
When $20\%$ of the in-band entries are randomly removed (second row), the missing elements disrupt spatial coherence within the band. This attenuates the sharp ridge structure and introduces spectral fluctuations, while the dominant orientation remains visible. The overall effect is a reduction in spectral concentration and increased diffusion of energy.
Adding uniformly distributed random noise across the entire matrix (third row) produces a similar effect. In both cases, the injected randomness contributes diffuse high-frequency energy, elevates spectral entropy, and weakens contrast in the dominant structures, partially diffusing spectral energy and masking signature patterns.


\subsection{Density-map Normalization.}
Due to the sensitivity of sparse spectra to random perturbations, we introduce a coarse \emph{density map} normalization to stabilize spectral signatures. Specifically, the matrix is partitioned into fixed-size tiles (e.g., $8\times8$), and each tile is replaced by the fraction of nonzero entries it contains. For a $1024\times1024$ matrix, this produces a $128\times128$ grayscale image that preserves global structural information while suppressing fine-scale randomness.
The FFT of the density map retains the dominant orientation and low-frequency energy concentration while attenuating high-frequency artifacts caused by missing entries and background noise. As shown in Figure~\ref{fig:sparsity-noise-fft} (right two columns), the density-map representation preserves the anti-diagonal ridge even under $20\%$ in-band sparsity and low-rate uniform noise, making the resulting spectral features more robust for detecting large-scale structures in real-world matrices.

However, this spatial compression reduces resolution. In particular, the effective band width becomes narrower after aggregation, and the magnitude spectrum of the density map may no longer exhibit the clear cross-shaped pattern observed in wide-band structures. Despite this loss of fine detail, the dominant directional and low-frequency signatures remain identifiable, allowing the underlying spatial pattern to be inferred from the coarse spectral representation. Furthermore, the spectral features from the density map and the original matrix can be complementary, where the former captures robust large-scale structure and the latter captures finer details. In Figure~\ref{fig:fft-importance}, we show that both types of features contribute to the spectral models' prediction accuracy.

\section{RQ2: Spectral analysis of real-world matrices}\label{sec:dataset-features}


To answer RQ2, this section applies the spectral descriptors defined in
Section~\ref{sec:background} to a large corpus of SuiteSparse
matrices, reports the observations, and demonstrates that unsupervised clustering in
each feature space recovers structurally meaningful groups. The resulting
clusters provide qualitative evidence that FFT features capture structural
diversity relevant to sparse-computation optimization, laying the foundation for the use of the features in optimization of sparse computing (exemplified in Section~\ref{sec:application}).

\subsection{Dataset}\label{sec:ori_dataset}
We extract 1{,}314 matrices from the SuiteSparse Matrix
Collection~\cite{davis2011suitesparse} and retain the sparsity pattern
only, converting every nonzero to one. This is because only positions of nonzeros matter for performance of sparse computation. Row counts range from 256 to ${\sim}2\times10^{6}$, and column counts from 256 to ${\sim}1.4\times10^{6}$.
The collection covers matrices from multiple categories: PDE discretizations, circuit simulations, optimization problems, and graph
adjacency matrices, providing a variety of structural patterns for explorations.

A large fraction of the SuiteSparse matrices in our dataset arise
from undirected graphs and are therefore symmetric; our method does not assume
or require matrix symmetry.

\begin{figure*}[htbp]
    \centering
    \begin{subfigure}[t]{0.32\linewidth}
        \centering
        \includegraphics[width=\linewidth]{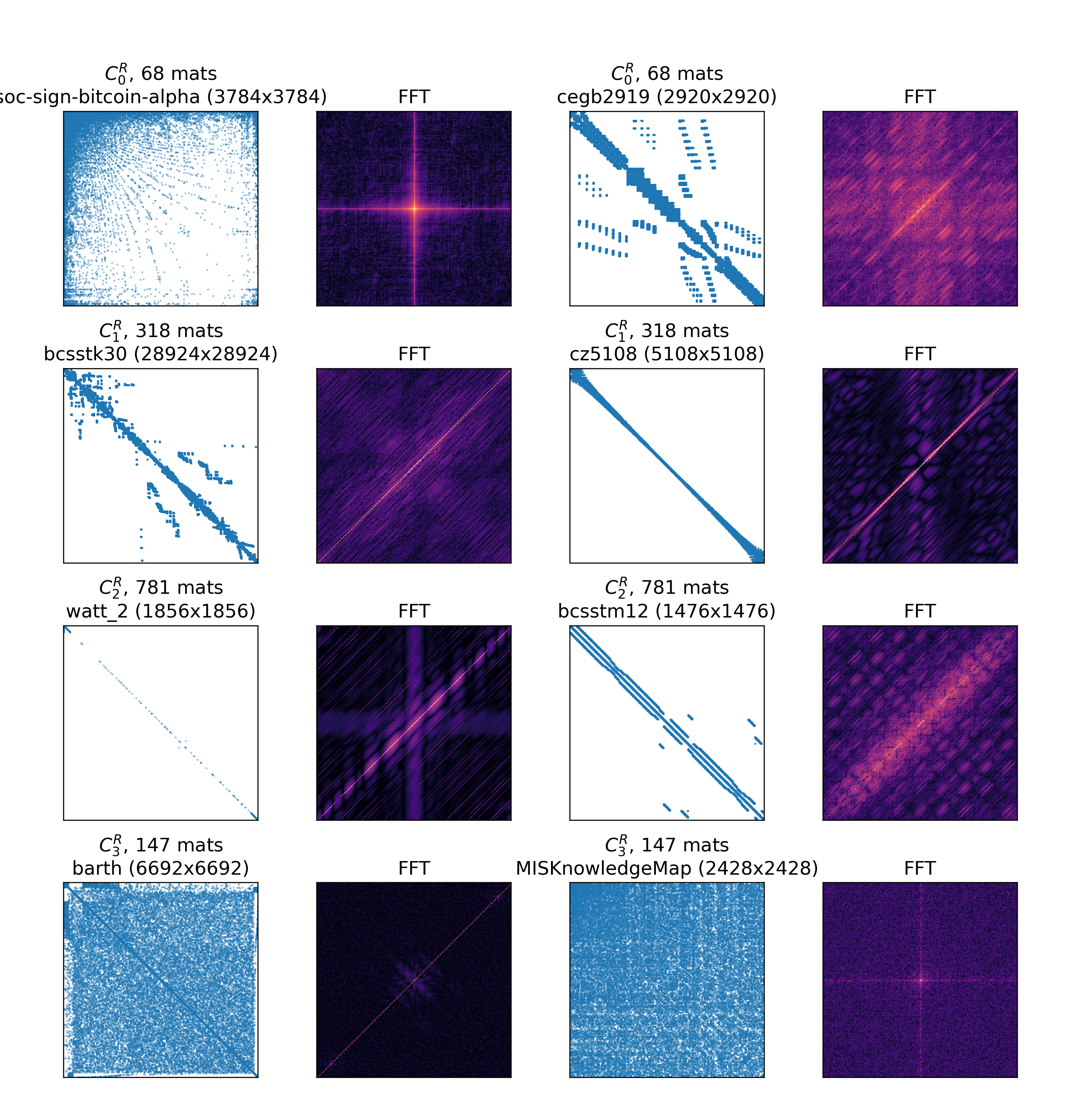}
        \caption{Radial energy}
        \label{fig:clustering-radial}
    \end{subfigure}
    \begin{subfigure}[t]{0.32\linewidth}
        \centering
        \includegraphics[width=\linewidth]{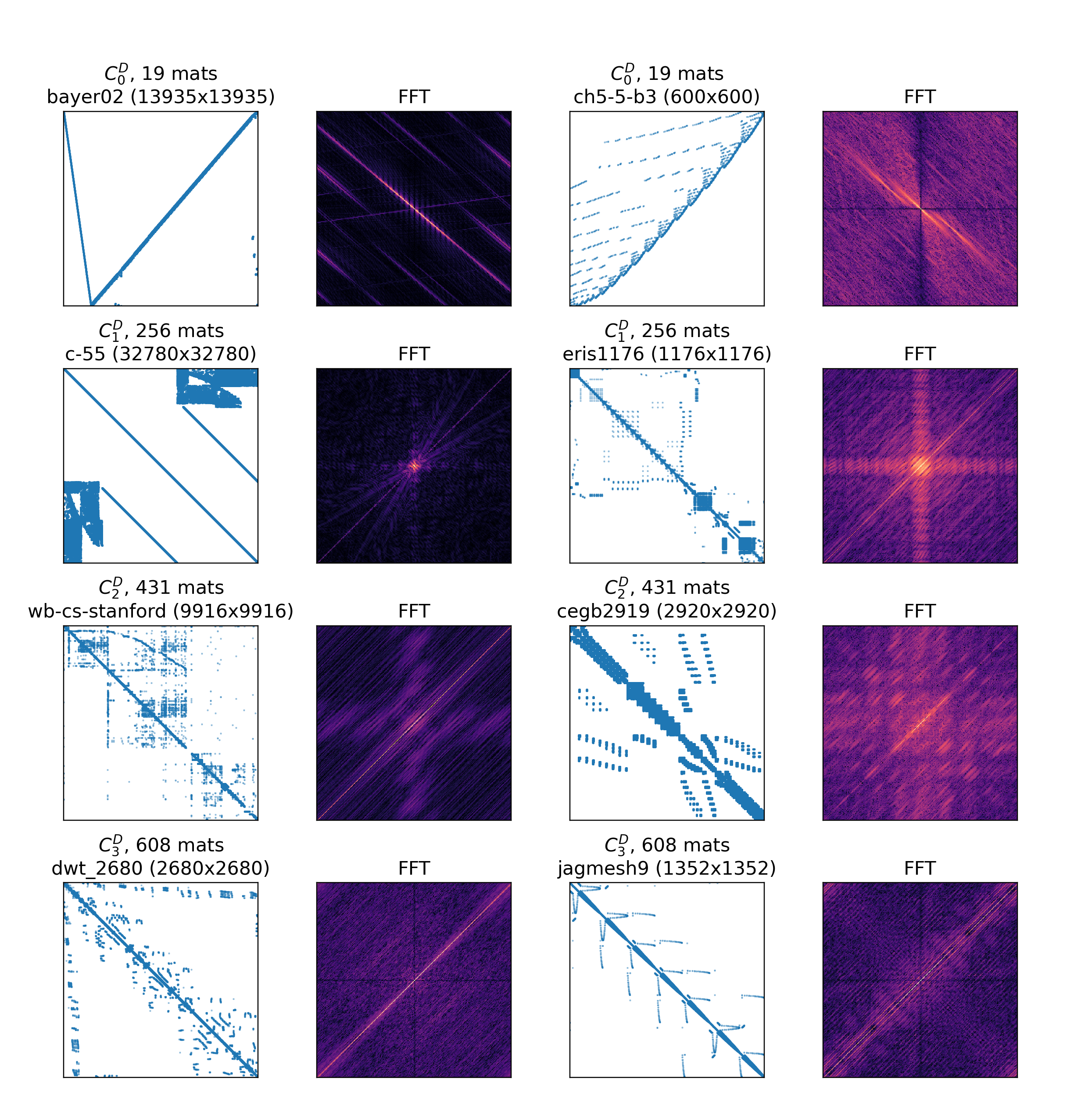}
        \caption{Directional energy}
        \label{fig:clustering-direction}
    \end{subfigure}
    \begin{subfigure}[t]{0.32\linewidth}
        \centering
        \includegraphics[width=\linewidth]{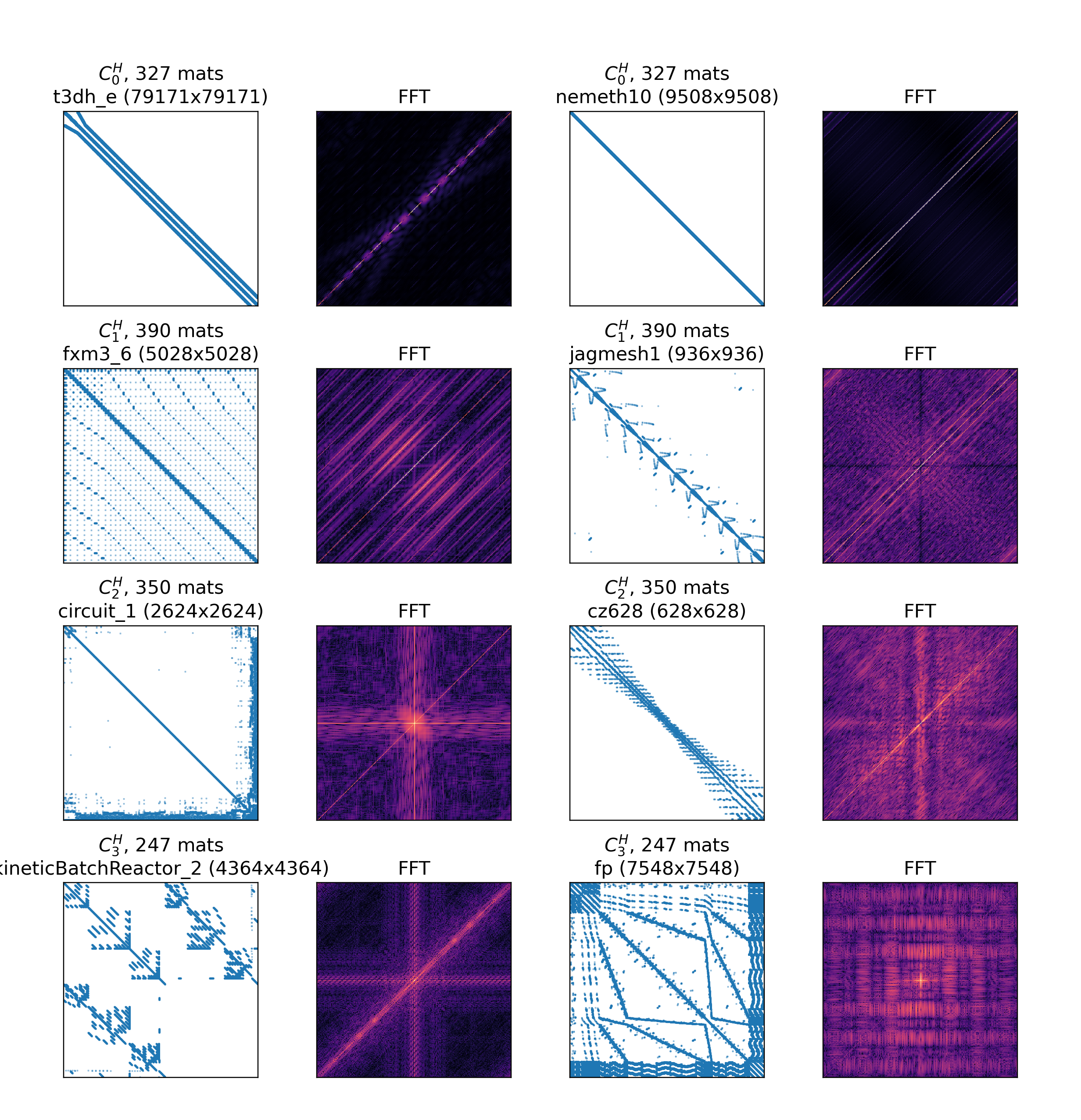}
        \caption{Spectral entropy}
        \label{fig:clustering-entropy}
    \end{subfigure}

    \caption{Clustering of matrices by radial energy (a), directional energy (b), and spectral entropy (c). $C_i^F$ denotes cluster $i$ for feature $F$. Each panel shows cluster representatives with their compressed $256{\times}256$ density maps and magnitude spectra.}
    \Description{Three grids of representative sparse matrices and spectra grouped by radial energy, directional energy, and entropy.}
    \label{fig:clustering-features}
\end{figure*}

We cluster each feature space independently with standard $k$-means algorithm. For clarity, we use ($k{=}4$) and
visualize one representative matrix per cluster (Figure~\ref{fig:clustering-features}).

\subsection{Observations}
Figure~\ref{fig:clustering-features} shows the clustering results for three classes of 
features (explained in Section~\ref{sec:background}). Each panel displays representative matrices from three
clusters alongside their density maps and FFT log-magnitude spectrum,
illustrating the structural diversity captured by each descriptor.

\textbf{Radial-energy clusters}
(Figure~\ref{fig:clustering-features}~\subref{fig:clustering-radial}).
Clusters are ordered by decreasing low-frequency energy fraction.
\textit{$C_0^R$} concentrates most of its magnitude in low frequencies. As illustrated, these matrices tend to be block-dominant, containing large, high-density blocks that produce a bright low-frequency core and a clear cross pattern in the spectra. They may benefit from block-based storage formats with appropriate block sizes.
\textit{$C_1^R$} and \textit{$C_2^R$} shift magnitude toward mid frequencies, indicating smaller structures and less regularity. The representatives show a dominant diagonal band, combined with scattered nonzero regions, which introduce mid- and high-frequency components beyond the dominant structure. These matrices may benefit from finer-grained block formats or hybrid schemes that adapt to both block-like regions and irregular sparsity.
\textit{$C_3^R$} emphasizes high-frequency energy and is dominated by irregular, noise-like patterns or very narrow structures.  Such matrices are more likely to benefit from general-purpose sparse formats designed to efficiently handle fine-grained sparsity.

\textbf{Directional-energy clusters}
(Figure~\ref{fig:clustering-features}~\subref{fig:clustering-direction}).
\textit{$C_0^D$} is a small region with patterns along the anti-diagonal direction. \textit{$C_1^D$} is characterized by strong concentration along the horizontal and vertical frequency axes, indicating pronounced row- or column-aligned structures such as block rows, block columns, or grid-like layouts. \textit{$C_2^D$} demonstrates a wider band on the horizontal and vertical frequency axes, indicating block structures with smaller size or contains higher intra-structure sparsity.
\textit{$C_3^D$} exhibits energy concentrated primarily along the diagonal orientation, corresponding to dominant diagonal bands. Storage schemes such as DIA may benefit from such near-diagonal patterns. Note that nearly half (46.27\%) of matrices fall into this community, indicating that diagonal patterns are common in the SuiteSparse dataset.

\textbf{Entropy clusters}
(Figure~\ref{fig:clustering-features}~\subref{fig:clustering-entropy}). Spectral entropy reflects the overall regularity of a matrix by measuring the concentration versus diffusion of spectral energy.
\textit{$C_0^H$} exhibits the lowest spectral entropy and groups highly regular matrices. As illustrated by the representative matrices, the spectrum shows strong concentration along specific orientations with limited diffusion elsewhere. When combined with directional features, low entropy helps confirm the presence of dominant block or diagonal structures with high spatial coherence.
\textit{$C_3^H$} has the highest entropy, indicating that spectral energy is distributed more uniformly across frequencies. This corresponds to matrices with more irregular sparsity patterns, where no single structure dominates.
\textit{$C_1^H,C_2^H$} demonstrates moderate entropy, representing matrices that contain a dominant structural component together with noticeable irregular regions. The representative examples show clear primary patterns, like the diagonal band in $jagmesh1$, accompanied by scattered or noisy components. Such mixtures are common in real-world sparse matrices, where structured regions coexist with irregular sparsity. In these cases, combining entropy with radial and directional features provides a more complete characterization.

\color{black}
\textbf{Combined FFT signals.} 
Across the previous three feature types, the clusters partition matrices along structurally meaningful axes: scale (radial energy), layout (directional energy), and overall regularity (spectral entropy).
Importantly, these axes are complementary rather than redundant. A matrix may belong to the ``low-frequency'' radial cluster, indicating large-scale structure, while simultaneously falling into the ``diagonal'' directional cluster, reflecting a wide near-diagonal band. Each descriptor captures a distinct structural aspect that cannot be inferred from the others alone. 

\begin{figure}[htbp]
	\centering
	\includegraphics[width=0.9\linewidth]{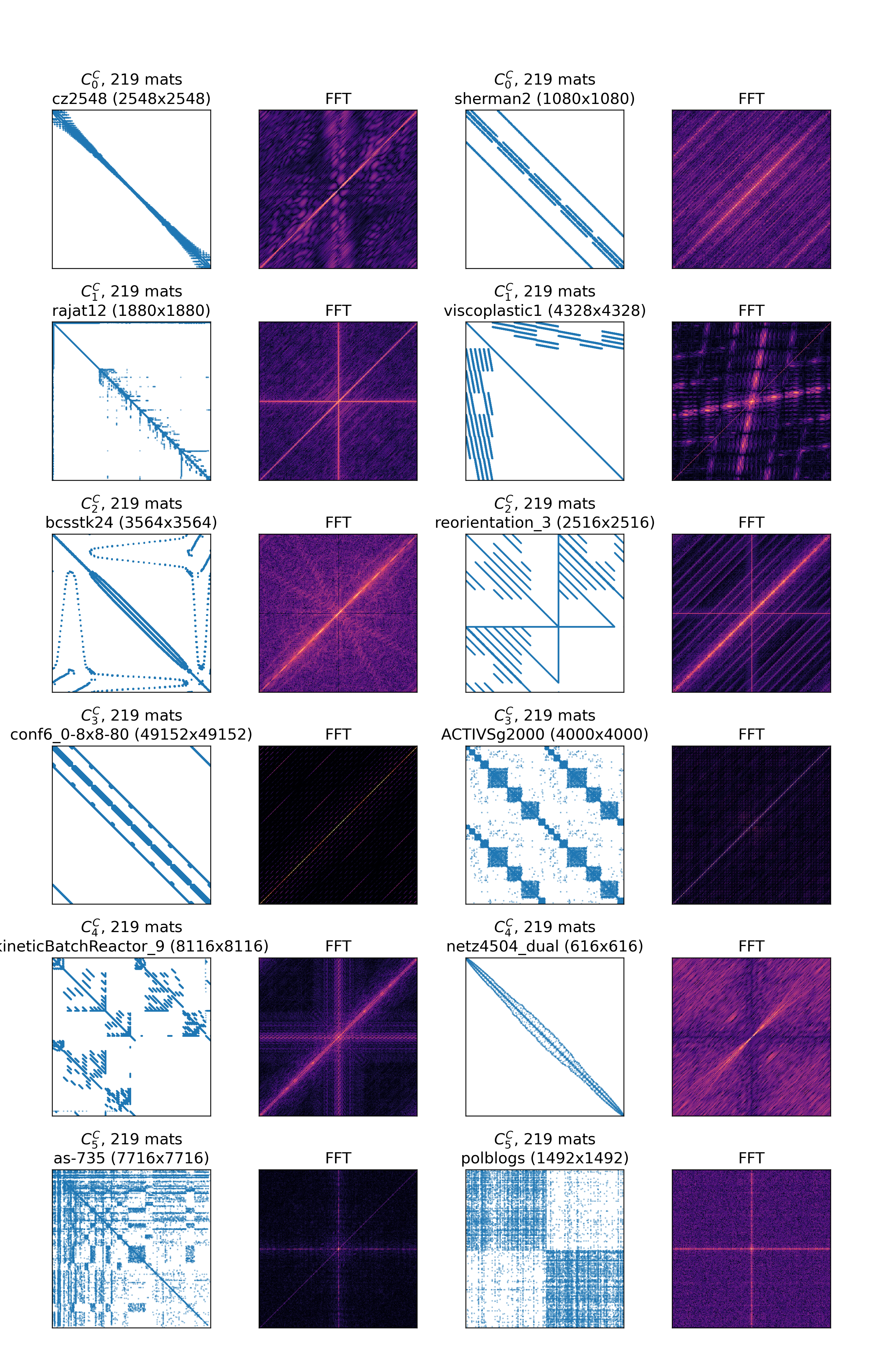}
	\Description{Representative sparse matrices and FFT spectra from balanced clustering over combined spectral features.}
	\caption{Clustering using combination of all features. Balanced K-means is applied to avoid tiny communities.}
	\label{fig:clustering-combined-clusters}
\end{figure}

\begin{figure}[htbp]
	\centering
	\includegraphics[width=0.8\linewidth]{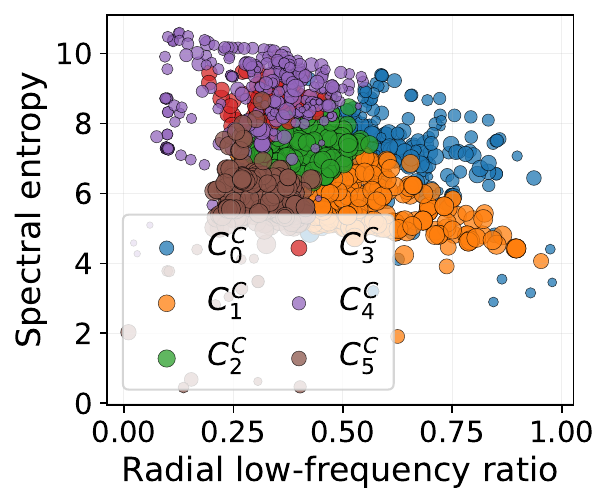}
	\Description{Scatterplot of low-frequency energy and entropy, with color denoting cluster and point size denoting diagonal directional strength.}
	\caption{Distribution of samples across clusters. The x-axis represents the low-frequency energy ratio, computed as the sum of the first four radial-energy bins; The y-axis denotes spectral entropy. The size of each point is proportional to the directional strength in the $45^\circ$ sector of the sparse matrix.}
	\label{fig:clustering-combined-scatter}
\end{figure}

Figure~\ref{fig:clustering-combined-clusters} shows the clustering results using the combination of all three feature types. To keep the feature dimensionality manageable, we sum up the first four radial bins from the Radial feature, indicating the ratio of low-frequency structure in the matrix. Additionally, we extract the anti-diagonal sector ($45^\circ$) from the Directional feature to quantify the degree of structural alignment with the diagonal. Figure~\ref{fig:clustering-combined-scatter} illustrates the sample distribution across the resulting clusters. We adopt balanced k-means to avoid assigning the majority of samples to only a few clusters.

The clusters from combined features exhibit clear structural patterns. For example, $C_2^C$ shows strong anti-diagonal energy together with moderate low-frequency concentration and entropy, suggesting matrices with mixed structures aligned along the diagonal direction. In contrast, $C_5^C$ is dominated by low-frequency energy but also has high entropy, indicating the presence of large-scale structures accompanied by substantial noise that disperses spectral energy.

These observations motivate the joint use of multiple spectral descriptors in downstream tasks such as kernel selection, where structural scale, orientation, and regularity influence format choice and block-size decisions in different ways.
\color{black}

\section{RQ3: Improving Sparse Computing through SpMV Kernel Selection}\label{sec:application}

To quantitatively study the usefulness of spectral analysis for improving sparse computing, we evaluate FFT-based spectral features for SpMV storage-format and kernel selection. SpMV can use a variety of sparse formats and corresponding GPU kernels, but no single choice is optimal for all matrices, and sparse patterns are critical to the selection~\cite{zhao2018bridging,yesil2023wise,zhou2019enabling,swann2024seer,shi2024dylaclass,zhang2026sparsex}. None of the prior work has used spectral features. We examine whether adding spectral features improves selection quality.
\subsection{Benchmark setup}

\subsubsection{Candidate Kernels}
Since different formats require different kernel functions, SpMV format selection is equal to kernel function selection. Because no single library contains a complete set of input formats for SpMV, we benchmark 14 SpMV kernels drawn from three GPU libraries:
Ginkgo~\cite{ginkgo-toms-2022} (CSR, COO, ELL, Hybrid, SELL-P with slice widths$\in\{16,32,64,128\}$, FBCSR-$8{\times}8$,
FBCSR-$16{\times}16$), TACO~\cite{kjolstad2017taco} (DIA), and
TileSpMV~\cite{niu2021tilespmv} (Fixed $16\times16$ tiles). These formats span the design space: CSR and COO are
general-purpose; ELL and SELL-P exploit uniform row lengths via padding and
slicing; DIA targets near-diagonal structure by storing only occupied
diagonals; Hybrid combines ELL and COO to handle mixed regularity;
FBCSR partition nonzeros into fixed-size tiles to exploit block
locality; TileSpMV further uses hybrid formats on different tiles to solve tile heterogeneity.

We primarily use Ginkgo library's implementation because we find that for the formats it supports, it generally outperforms other libraries such as TACO and cuSPARSE. And benchmarking kernel implementations from the same library generally better reflects the performance gaps caused by the sparse patterns. We conducted the measurements on NVIDIA RTX 4090 GPU.

\begin{figure}[htbp]
	\centering
	\includegraphics[width=0.7\linewidth]{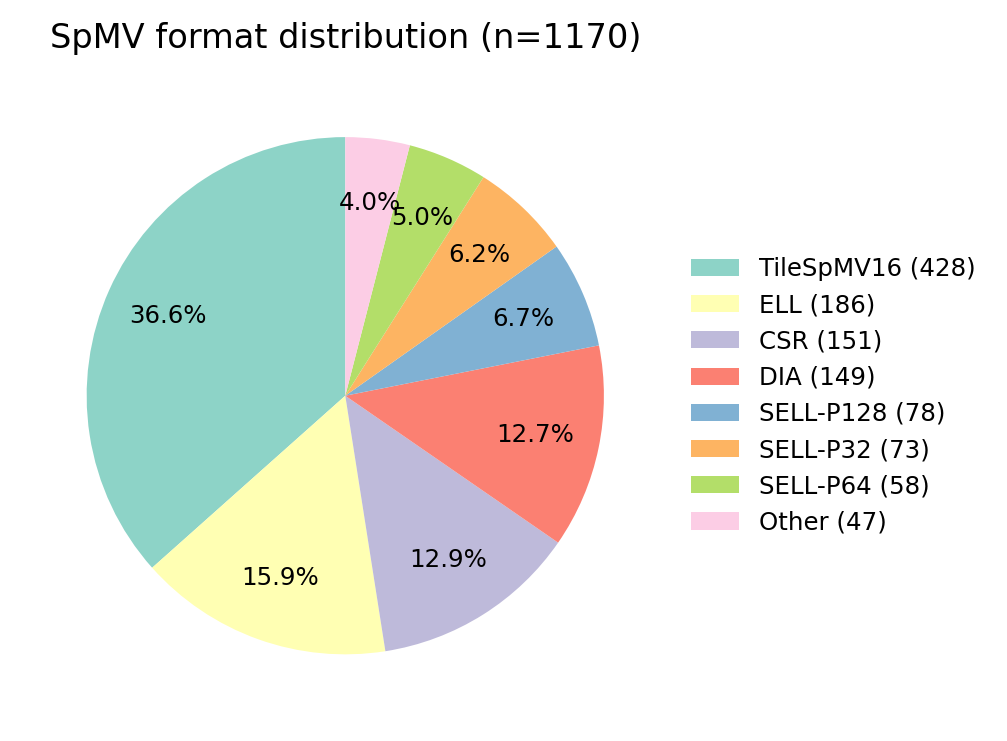}
	\Description{Pie chart showing the best-kernel distribution for the SpMV benchmark suite.}
	\caption{Distribution of the best kernels for the 1,170 SuiteSparse SpMV
		matrices. Only classes with $\geq$5 samples are retained for training.}
	\label{fig:kernel-dist}
\end{figure}

\subsubsection{Datasets} We use two matrices sets.
\noindent{\bf SuiteSparse.}
The first is the set of commonly used SuiteSparse~\cite{davis2011suitesparse} matrices mentioned in Section~\ref{sec:dataset-features}. Each matrix is timed with all 14 candidate kernels; Figure~\ref{fig:kernel-dist} shows the distribution of the fastest kernels on the matrices.
\noindent{\bf LLM-Matrices.}
The second is a more recent collection, {\em LLM-Matrices}~\cite{prunedllmmatrices2026}. It contains 1056 matrices from real-world Large Language Models (LLM), Qwen2.5-3B~\cite{yang2024qwen25}, OPT-2.7B~\cite{zhang2022opt},
Llama-2-7B~\cite{touvron2023llama2}, and Mistral-7B~\cite{jiang2023mistral7b}. These models were pruned, an important optimization for efficiency~\cite{frantar2023sparsegpt,sun2024wanda}. The pruning uses magnitude-based pruning~\cite{han2015learning} and WANDA~\cite{sun2024wanda}; it focuses on \texttt{q\_proj} and
\texttt{o\_proj}/\texttt{out\_proj} attention layers. The sparsities are 80\% and 90\%, and sizes range from $2048{\times}2048$ to $4096{\times}4096$. SpMV is the primary operation in batch-one autoregressive decoding of LLM.


\subsection{Feature sets and model}

We frame kernel selection as multi-class classification and compare four feature
sets to isolate the contribution of FFT features. 

\textbf{Base features.} The baseline uses four statistics: matrix dimensions,
the number of nonzeros, and density. These features capture size and workload
volume but provide no structural information about the nonzero pattern.

\textbf{Spatial features.}
We use the WISE feature set~\cite{yesil2023wise} as the representative of the spatial feature set in SOTA. WISE remains the most comprehensive
hand-engineered characterization of sparse-matrix structure in the ML-based
kernel selection literature. Later works either focus on different aspects~\cite{sun2023manet,shi2025automatic} or use simpler features for efficiency~\cite{swann2024seer,shi2024dylaclass,zhang2026sparsex}. To the best of our knowledge, there is no feature set in prior work that exceeds WISE's coverage. 

The feature set characterizes sparse matrices along three dimensions: size, nonzero skew, and nonzero locality.

\begin{itemize}
    \item \textit{Size features:}
    Similar to the Base features, capturing the problem scale and workload.

    \item \textit{Nonzero skew features:}
    Summary statistics of nonzero distributions over rows (R) and columns(C), including $\mu$, $\sigma$, $\sigma^2$, min/max, Gini coefficient ($G$), $p$-ratio ($P$), and non-empty counts ($ne$). These features affect row scheduling and memory access patterns.

    \item \textit{Nonzero locality features:}
    Statistics derived from distributions of tiles (T), row blocks (RB), and column blocks (CB), along with intra-tile statistics (e.g. number of unique rows containing nonzeros in each tile), capturing memory access locality across cache levels. A matrix is broken into $128\times128$ tiles for our experiments.
\end{itemize}
Further details can be found in Section 4.2 of WISE paper~\cite{yesil2023wise}. We denote the model using WISE features as \emph{Spatial}.


\textbf{Spectral features (FFT).} We extract the spectral features at three density-map scales:
\textit{full}, $1024{\times}1024$, and $256{\times}256$. For the full scale,
we apply FFT on the original matrix map unless the input is too large $(\geq 10k)$, in
which case we use capped compression to keep FFT cost bounded. The two resized
scales provide medium/coarse views of global structure and improve
cross-matrix comparability.

At each scale, we extract the same nine FFT feature families from the shifted, log-scaled magnitude spectrum:
\begin{itemize}
	\item \textbf{Radial$_i$/Direction$_i$/Entropy:} basic spectral features described in Section~\ref{sec:background}. We use 16 bins for Radial features.
	\item \textbf{Radial$_{\mathrm{Low/Med/High\text{-}freq}}$:} similar to radial but coarse low/mid/high frequency-band energy ratios.
	\item \textbf{Quadrant$_i$:} spectral energy ratio across the quadrants.
	\item \textbf{Flatness:} measures the uniformity of spectral energy across frequencies. Higher flatness indicates a more even, noise-like distribution, while lower flatness indicates energy concentrated in a few frequency components.
	\item \textbf{Bandwidth:} dispersion of spectral energy around the centroid; larger values indicate broader frequency spread, while smaller values indicate tighter concentration.
	\item \textbf{Energy\_Avg/Energy\_Std:} average and standard deviation of the log-scaled magnitude spectrum.
\end{itemize}
These features have 33 dimensions per scale. In practice, because some features exhibit internal correlations (e.g., directional energy ratios sum to one) and certain spectral features capture overlapping information (e.g., energy and radial descriptors), we first use an XGBoost model to select the most important FFT features, retaining only the top 20 features for SpMV. Figure~\ref{fig:fft-importance} shows the top-10 FFT features. Note that if more matrices with diverse structures are added to the dataset, incorporating additional FFT features may prove beneficial.
These FFT descriptors are the \emph{spectral} features. Models using only spectral features are denoted
\emph{Spectral}; and models using both spatial and spectral features are \emph{Spatial+Spectral}.

\textbf{Model and training.} We employ XGBoost~\cite{chen2016xgboost} with multi-class softmax classification for all experiments. The XGBoost model has 480 trees, with maximum depth of 6, learning rate of 0.02, L1 and L2 regularization coefficients of 1.0, and subsampling rates of 0.8.
For SuiteSparse, we perform 5-fold cross-validation by partitioning matrices,
so a matrix appears in only one fold. The pruned-LLM experiment instead uses the
leave-one-model-out protocol described in Section~\ref{sec:llm-application}.

\subsection{Metrics \& overhead}
We focus on two metrics. (i) Selection accuracy: the percentage of matrices on which the selected kernel is indeed the fastest. Because of the fluctuations in the timing, we report the selection accuracy in two cases, with $r=0$ or $5$ tolerance (i.e., a prediction is considered correct if the predicted kernel's runtime is within $r\%$ of the oracle (fastest) runtime). (ii) Speedups: the SpMV running time (avg of 50 repeats) of the default kernel divided by that of the selected kernel. 

Note that the times in feature extraction and prediction are not included in the speedup calculation. It is because the primary use cases of kernel selection, as prior work~\cite{zhao2018bridging,yesil2023wise,swann2024seer,shi2024dylaclass,zhang2026sparsex} focuses on, is where the same sparse matrix (e.g., a graph or an LLM model) is used for many times. Prior studies report that the overhead of kernel selection is typically equal to hundreds or thousands of iterations of SpMV. Adding spectral features add some extra time, ranging from tens to hundreds of iterations. But either way, the selection will benefit a much greater number of post-deployment iterations, and hence the overhead is mainly not counted in evaluating the efficacy of kernel selection.


\subsection{Results}

Our experiments show that the spectral features complement the spatial features, able to consistently boost the prediction accuracy of kernel selection, even on SuiteSparse, a dataset heavily studied in prior work. Its improvements are significantly more pronounced on LLM-matrices, a newer set from the real world. 

\subsubsection{SuiteSparse Matrices}
\color{black}
\begin{table}[htbp]
	\caption{SpMV format selection accuracy (strict and 5\%
		tolerance). Adding spectral features to the spatial baseline improves
		tolerant accuracy by 3.6 percentage points.}
	\label{tab:spmv}
	\centering
	\begin{tabular}{lccr}
		\toprule
		Feature set      & Tolerant Acc.\  & Strict Acc.\   & \#Feat \\
		\midrule
		Base             & 0.615           & 0.564          & 4      \\
		Spectral (FFT)   & 0.723           & 0.650          & 24     \\
		Spatial (WISE)   & 0.757           & 0.676          & 84     \\
		Spatial+Spectral & \textbf{0.793}  & \textbf{0.701} & 104    \\
		\bottomrule
	\end{tabular}
\end{table}

\begin{figure*}[htbp]
	\centering
	\begin{minipage}[htbp]{0.4\linewidth}
		\centering
		\includegraphics[width=\linewidth]{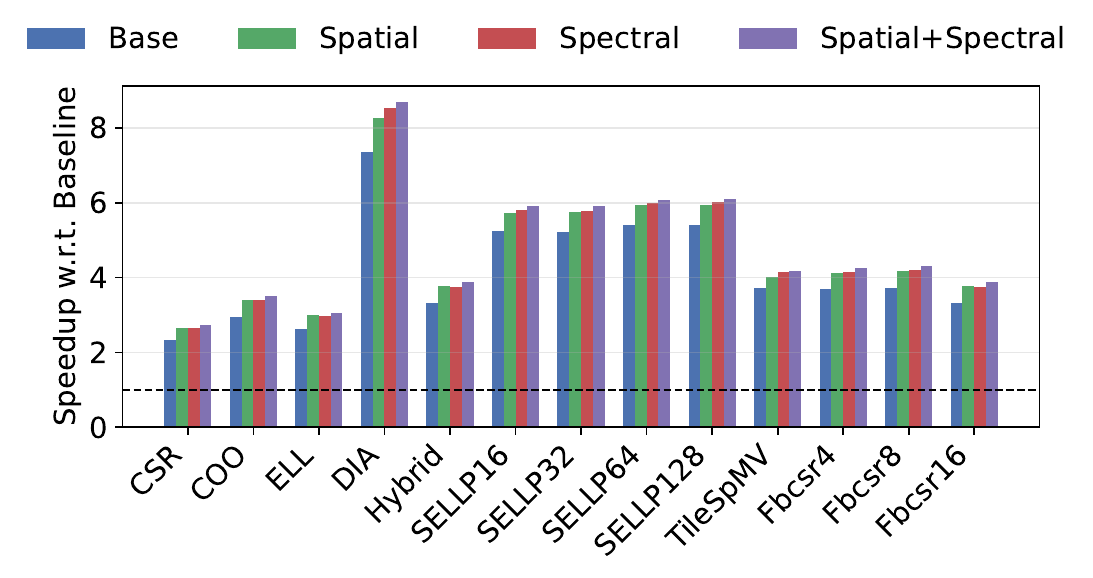}
	\end{minipage}\hfill
	\begin{minipage}[htbp]{0.5\linewidth}
		\centering
		\includegraphics[width=\linewidth]{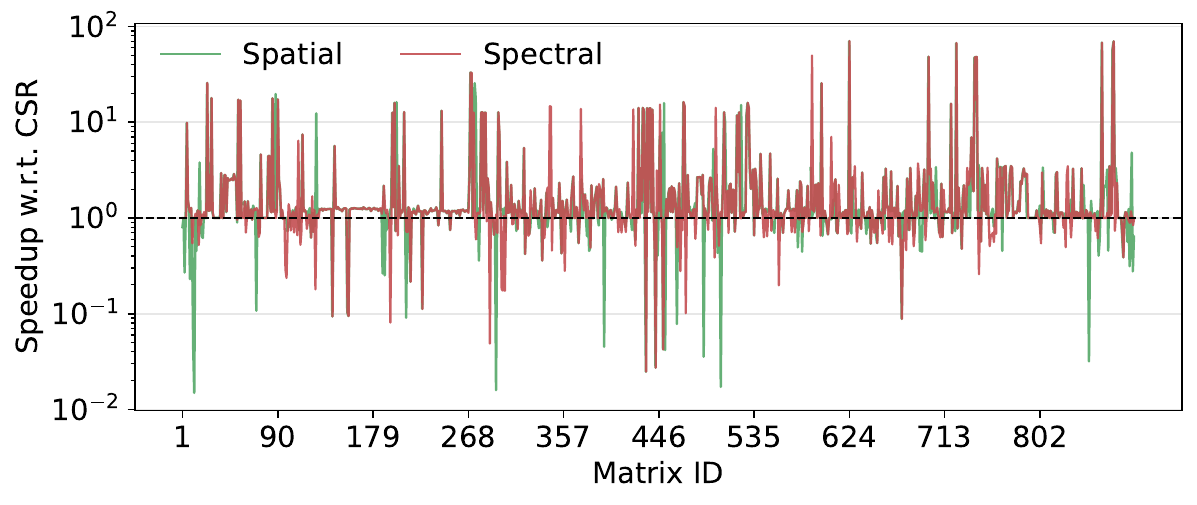}
	\end{minipage}
	\caption{SpMV speedup of model-predicted formats. Left: speedup versus each
		fixed baseline kernel ($y{=}1$ indicates parity). Right: per-matrix
		speedup versus CSR, sorted in ascending order of low-frequency energy ratio.}
	\Description{A bar chart of model-selected SpMV speedups and a per-matrix speedup plot sorted by spectral energy.}
	\label{fig:spmv-speedup}
\end{figure*}

Table~\ref{tab:spmv} reports the performance of the four models on the SpMV dataset. Both strict accuracy and 5\% tolerance accuracy are reported. Unless otherwise specified, we refer to the tolerance-based accuracy.

Using only basic matrix size and sparsity features achieves 61.5\% accuracy. Introducing FFT-based spectral descriptors increases accuracy to 72.3\%, demonstrating that spectral signals capture structural information critical to sparse kernel performance. The Spatial model achieves 75.7\% accuracy, slightly outperforming the Spectral model. This advantage stems from features that directly reflect workload balance, such as row-wise nonzero distributions. For example, matrices with extremely long rows can make formats like ELL impractical due to excessive padding overhead---patterns that are not easily inferred from the spectral alone.

The Spatial+Spectral model achieves the highest accuracy at 79.3\%, indicating that integrating spatial and spectral features yields the most reliable predictions. Spatial features alone cannot fully capture global patterns such as diagonal bands favoring DIA or hybrid structural combinations benefiting kernels like TileSpMV. Figure~\ref{fig:case-study-fft} further illustrates how spectral features improve predictive decisions.

Figure~\ref{fig:spmv-speedup} shows that improved prediction accuracy translates into runtime gains. What is especially notable is that Spatial+Spectral achieves the highest speedup over fixed baselines {\em across all the formats/kernels} (Figure~\ref{fig:spmv-speedup} left), despite that SuiteSparse was already heavily studied in prior kernel selection work. 
Although the Spectral model achieves slightly lower accuracy than the Spatial model, the achieved speedups using the predicted formats of the two models are comparable (2.65$\times$ vs\ 2.66$\times$ over CSR). As shown in the right panel of Figure~\ref{fig:spmv-speedup}, even when the Spectral model fails to predict the optimal kernel exactly, it tends to avoid severe slowdowns caused by poor choices. 


\begin{table}[htbp]
  \caption{SpMV format selection results for pruned LLM decode. Acc denotes tolerant accuracy. Speedup
  compares Spatial+Spectral model's kernel runtime against Spatial model.}
  \label{tab:llm-finegrained}
  \centering
  \scriptsize
  \setlength{\tabcolsep}{2.5pt}
  \begin{tabular}{@{}llccc@{}}
    \toprule
    Pruning Method & Sparsity & Spatial Acc. & Spatial+Spectral Acc. & Speedup \\
    \midrule
    Magnitude & 80\% & 73.18\% & \textbf{83.64\%} & \textbf{1.110$\times$} \\
    Magnitude & 90\% & 82.95\% & \textbf{86.52\%} & \textbf{1.035$\times$} \\
    \addlinespace
    WANDA & 80\% & 82.20\% & \textbf{95.53\%} & \textbf{1.245$\times$} \\
    WANDA & 90\% & 68.79\% & \textbf{75.15\%} & \textbf{1.073$\times$} \\
    \midrule
    Overall & 80/90\% & 76.78\% & \textbf{85.21\%} & \textbf{1.087$\times$} \\
    \bottomrule
  \end{tabular}
\end{table}

\subsubsection{LLM-matrices}
\label{sec:llm-application}
As shown in Table~\ref{tab:llm-finegrained}, across the four sparsity and pruning methods, Spatial+Spectral raises accuracy from 76.78\% to 85.21\% and achieves 1.087x speedup. 
 Notably, spectral features yield substantially larger gains on WANDA-pruned matrices than on magnitude-pruned ones. Unlike magnitude pruning, WANDA scores each weight using both its magnitude and the norm of its corresponding input activations, making the resulting pruning mask sensitive to channel-wise activation heterogeneity. This produces a more nonuniform distribution of nonzeros across input channels, whose global structure is better captured by spectral features. Also, adding spectral features better boosts model performance at lower sparsity levels, where patterns like dense blocks are more likely to exist.

To connect the sparse kernel result to application latency, we measure a
batch-one, one-token decode with a fixed 128-token KV cache. With WANDA pruning and 90\% sparsity setting, the sparse projections occupy 8.63\% of Qwen's decode time and 14.45\% of OPT's.
The Spatial+Spectral model achieves $4.37$\% and $7.47$\% end-to-end speedup on the two LLM models compared to using GEMM, respectively.
\color{black}

\begin{figure}[htbp]
	\centering
	\includegraphics[width=\linewidth]{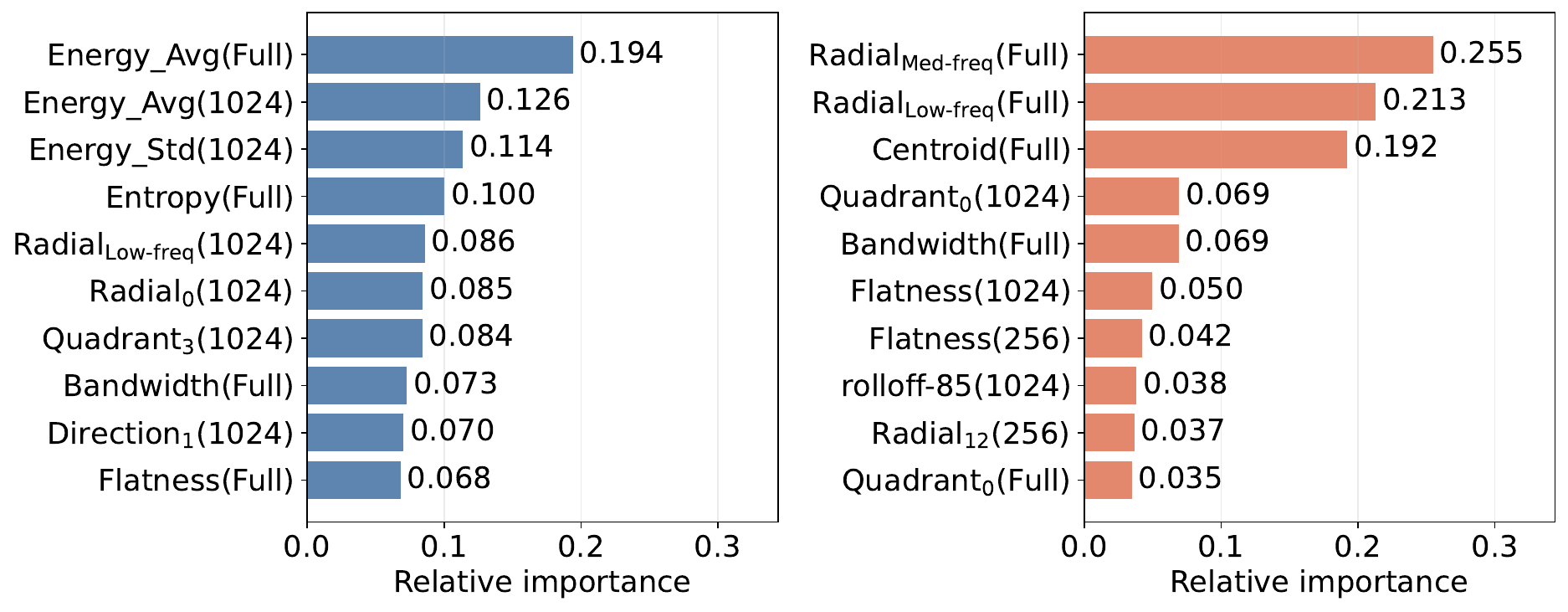}
	\Description{Two side-by-side ranked bar charts of the ten most important FFT features for SuiteSparse and pruned-LLM SpMV Spectral predictors.}
	\caption{Top-10 FFT features for SpMV Spectral models. Left: SuiteSparse; right: LLM-matrices.}
	\label{fig:fft-importance}
\end{figure}

\subsection{Spectral Feature Importance Analysis}
To analyze which spectral signals are most influential for SpMV, we extract feature importance scores from the Spectral XGBoost models. Figure~\ref{fig:fft-importance} presents the top-10 FFT-based features for SuiteSparse and pruned-LLM SpMV.

The most important feature is \texttt{Energy\_Avg} on the full-scale spectrum, reflecting the average structural density in the frequency domain. This is followed by energy-based indicators derived from the $1024\times1024$ density map. Spectral entropy further measures the uniformity of the energy distribution, while radial and bandwidth features emphasize the proportion of low-frequency energy, corresponding to large-scale spatial coherence. Directional, quadrant, and flatness features highlight orientation patterns and dominant block structures or diagonal bands.
For pruned-LLM SpMV, full-scale mid- and low-frequency radial energy and
the spectral centroid are most important, followed by quadrant, bandwidth, and
flatness descriptors. This ranking shows that pruned LLM layers tend to contain more large sparse structures, which can be observed from the Radial distribution.

\begin{figure}[htbp]
	\color{blue}
	\centering
	\includegraphics[width=\linewidth]{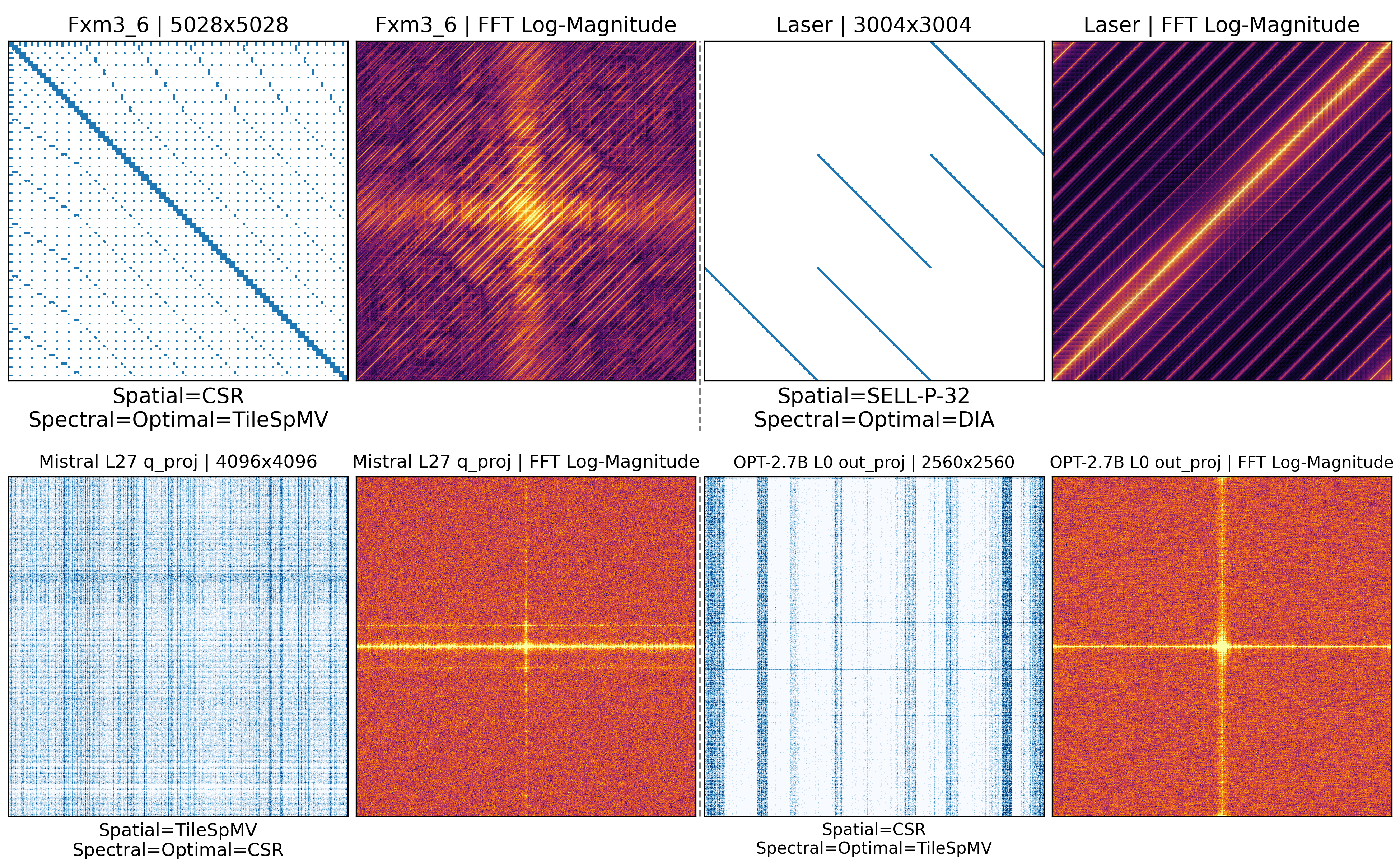}
	\Description{Four sparse-matrix and FFT pairs arranged two per row: two SuiteSparse SpMV cases followed by two pruned-LLM SpMV cases.}
	\caption{Four case-study examples in the same sparse-matrix/FFT format where
		Spectral models select the fastest kernel while Spatial models miss.}
	\label{fig:case-study-fft}
\end{figure}

\subsection{Case Study: Effectiveness of Spectral Signals}
To further showcase the relationship between FFT-based structural signals and optimal computation kernels, Figure~\ref{fig:case-study-fft} presents four representative cases in which spatial features alone fail to predict the optimal kernel, whereas spectral features provide clear guidance.

For the SpMV task on \texttt{fxm3\_6}, the optimal choice is the TileSpMV kernel, yet the spatial model predicts CSR. From the spatial perspective, the nonzeros are evenly distributed in rows and columns, favoring fine-grained CSR format. However, the FFT spectrum reveals a wide diagonal band manifested as cross-shaped energy concentration near the horizontal and vertical frequency axes. In addition, substantial high-frequency components indicate the presence of irregular structures, corresponding to dotted residuals in the original matrix. The hybrid tile-based design of TileSpMV effectively captures the dominant diagonal structure while accommodating irregular components, explaining its superior performance.
As for \texttt{laser}, the FFT spectrum shows a strong anti-diagonal ridge, clearly indicating a dominant diagonal structure. This pattern aligns well with the DIA format, which is efficient when nonzeros are concentrated along a small number of diagonals. In contrast, the spatial model lacks explicit features to capture diagonal alignment. While spatial descriptors could be extended (e.g., by incorporating diagonal density distributions), FFT features naturally expose such orientation information without requiring additional handcrafted metrics.

The bottom row shows two pruned LLM weights. For the Mistral-7B
matrix, Spatial predicts TileSpMV, whereas Spatial+Spectral recovers the oracle
CSR kernels, reducing runtime from 17.803 to
7.196~$\mu$s (2.474$\times$). Its spectrum is relatively diffused, indicating irregular sparsity. For the OPT-2.7B matrix, Spatial predicts CSR,
whereas Spatial+Spectral recovers the oracle TileSpMV kernel,
reducing runtime from 8.202 to 1.616~$\mu$s (5.076$\times$). The clear low-radial concentration indicates large blocks, and the diffused spectrum entropy shows that there are various patterns, making TileSpMV the oracle kernel here.
\section{Related Work}\label{sec:related}

Earlier sections have already covered sparse formats and kernel selection, which are skipped here.

Spectral analysis has been widely used in many domains, {\bf but to the best of our knowledge, it has not been introduced into sparse matrix computing.} 

The area that is closest to our focus and has seen the use of spectral analysis is graph computing~\cite{von2007tutorial,spielman2012spectral}. 
Graph Fourier
Transform (GFT) represents signals on graph vertices in a spectral domain
defined by the eigenbasis of graph Laplacian or adjacency operators~\cite{shuman2013emerging}.
Building on this spectral foundation, convolutional neural networks on graphs
were proposed by defining convolution in the graph spectral domain~\cite{bruna2013spectral}.
This formulation was later simplified by the Graph Convolutional Network (GCN)
as a first-order approximation of spectral graph convolution~\cite{kipf2016semi}.
GCNs have since been widely used in graph-related tasks such as node
classification~\cite{zhao2019t} and community detection~\cite{he2021community}.

Graph spectral analysis is related to our work because some sparse matrices
originate from graph adjacency matrices. Methods like GCN primarily focus
on graph signal processing through spectral filtering defined in the eigenbasis
of the graph Laplacian. In contrast, our approach applies a 2-D FFT directly to
the nonzero pattern of a sparse matrix by treating it as an image, which uses a
fixed two-dimensional Fourier basis defined on a regular grid, making it
fundamentally different from previous frequency-domain analyses on graphs. A more concrete example is that some graph reordering algorithms permute graph nodes to manipulate sparse patterns and accelerate sparse computations~\cite{chen2025accelerating}. Such permutations preserve many graph spectral properties (graph invariants), such as the Laplacian eigenvalues. In contrast, applying a 2-D FFT to the sparse matrix treated as an image is highly sensitive to these permutations and can reflect the resulting pattern changes directly in the frequency domain.

\section{Conclusion}
By linking performance-relevant spatial structures to their corresponding spectral signatures, this work presents the first framework that bridges FFT-based spectral signals with sparse computation optimization.
Across SuiteSparse and pruned-LLM SpMV kernel selection, FFT-derived features provide a compact and informative representation that complements traditional spatial metrics. This combination improves predictor accuracy and selected-kernel runtime: on pruned LLM decoding, adding spectral features yields 1.035--1.245$\times$ kernel speedups. This demonstrates that spectral analysis is a practical and effective tool for sparse matrix characterization and performance optimization.
\clearpage


\bibliographystyle{ACM-Reference-Format}
\bibliography{refs}

\newcommand{\preservedLLMGrid}{%
\noindent\textbf{Post-hoc sensitivity analysis (not confirmatory evidence).}
The following 704-matrix subsection and Table~\ref{tab:llm-pruning-spmv} are
preserved for sensitivity context; the complete confirmatory protocol appears
in Section~\ref{sec:llm-application}.

We further evaluate the strongest repeat-validated LLM case.  Starting from
Qwen2.5-3B, OPT-2.7B, Llama-2-7B, and Mistral-7B, we sample eight transformer
blocks per model and the square \texttt{q\_proj}/\texttt{o\_proj} attention
projections.  Applying 33 magnitude, activation-aware, $2{:}4$, row/channel,
and square-block pruning configurations produces 2,112 distinct sparse weight
patterns.  The candidate set is the complete 14-kernel SpMV suite described
above, and $N{=}1$ models token-by-token decode with a single active sequence.
This setting follows the post-training sparse-LLM use case studied in
SparseGPT~\cite{frantar2023sparsegpt}, while our evaluation focuses on kernel
selection rather than language-model quality.

We report only the strongest setting that remains positive under independent
repeat timing: a model-balanced 704-matrix subset containing unstructured
magnitude, activation-aware unstructured, and block-16 pruning.  We use
leave-one-model-out evaluation---training on three model families and testing
on the fourth---and keep the selector predictions fixed during timing
validation.  Every matrix--kernel pair is measured in three independent runs,
and the median runtime is used to reconstruct the oracle and 5\%-tolerant
labels.

\noindent\textit{The following table is a post-hoc sensitivity result, not
confirmatory evidence.}
\begin{table}[htbp]
	\color{blue}
	\renewcommand{\thetable}{\Roman{table}-legacy}
	\caption{Best repeat-validated LLM decode ($N{=}1$) case over 704 pruning
		matrices and 14 SpMV kernels (mean over five selector seeds; runtimes are
		sums over the validation matrices).}
	\label{tab:llm-pruning-spmv}
	\centering
	\footnotesize
	\begin{tabular}{lccc}
		\toprule
		Feature set & Tolerant Acc. & Selected Time (ms) & Speedup \\
		\midrule
		Spatial          & 84.26\% & 4.0376 & 1.0000$\times$ \\
		Spatial+Spectral & \textbf{92.24\%} & \textbf{3.8218} & \textbf{1.0565$\times$} \\
		\midrule
		\multicolumn{4}{l}{\textit{Complete-task timing: + full $2{:}4$ family (832)}}\\
		Spatial          & 86.63\% & 5.8287 & 1.0000$\times$ \\
		Spatial+Spectral & \textbf{93.56\%} & \textbf{5.6105} & \textbf{1.0389$\times$} \\
		\bottomrule
	\end{tabular}
\end{table}
\addtocounter{table}{-1}
\addtocounter{table}{1} 

Adding FFT features therefore improves tolerant accuracy by 7.98 percentage
points and reduces aggregate selected-kernel time by 5.34\%.  The speedup is
positive for all five selector seeds (1.0532$\times$--1.0593$\times$).  Across
9,856 matrix--kernel groups, the median relative timing range is 0.159\%, well
below the observed aggregate gain.  This result is a kernel-selection speedup
for sparse attention projections; it does not include dense layers, attention,
KV-cache management, framework overhead, or pruning-quality effects, and
therefore is not presented as end-to-end LLM latency.

\noindent\textit{Coverage stress test.}  The added 128 magnitude- and
activation-weighted $2{:}4$ matrices form a complete model-balanced family and
are neutral by themselves: both feature arms attain 100\% tolerant accuracy
and identical selected time.  The 832-matrix panel uses the complete-task
fixed predictions and three-round medians from
Section~\ref{sec:llm-application}; it broadens sparsity-pattern
coverage but supplies no independent FFT gain.  Its speedup remains positive
for every selector seed (1.0327$\times$--1.0438$\times$), while the held-out
Llama boundary remains negative at 0.9913$\times$.
\color{black}

\color{blue}
\noindent\textbf{Scope clarification.}  Table~\ref{tab:llm-pruning-spmv} is a
post-hoc sensitivity analysis of a favorable 704-matrix slice and is not used
as the headline transfer result.  The confirmatory analysis in Section~\ref{sec:llm-application} instead
retains all 2,112 matrices under a protocol fixed before the final timing
rounds; consequently, we do not use the slice's 1.0565$\times$ speedup to
support the main claim.

\noindent\textbf{Balanced pruning-grid sensitivity analysis}\\
\textbf{(exploratory).}
To expose rather than select among structures within the predeclared
global/square-block grid, we complement the asymmetric 704-matrix slice with a
complete factorial over two scores
(magnitude and activation-weighted), six granularities (global and square
blocks of size 4, 8, 16, 32, and 64), and three sparsities (70\%, 80\%, and
90\%).  Applied to the same 64 model/layer/projection sources, this produces
36 configurations and 2,304 sparse matrices, with exactly 18 configurations
per score family and 192 matrices in every granularity--score row.  The
activation-weighted score is $|W|$ times input-channel activation RMS from the
fixed 16-prompt calibration set; thresholds are matrix-global at the tensor or
block granularity.  We therefore use the term \emph{activation-weighted}, not
canonical Wanda, and make no language-model quality claim.

We retrain the unchanged five-seed leave-one-model-out selectors on the full
balanced collection and evaluate the same 14 SpMV candidates at decode
$N{=}1$.  The eleven newly required activation-weighted block configurations
(704 matrices) have exact three-round coverage; every TileSpMV check passes,
and each matrix--kernel label uses its three-round median.  Table~\ref{tab:llm-pruning-balanced-grid}
reports every granularity--score combination; the deployment-distinct $2{:}4$
family is not mixed into this factorial.

For timing provenance, we reuse the locked three-round medians for the 1,600
existing matrices and collect three clean rounds for the 704 newly generated
matrices; both feature arms use identical per-matrix timing labels.  Across the
9,856 new matrix--kernel groups, the relative timing range has median 0.132\%
and 95th percentile 1.483\%, but a long-tailed maximum of 137.0\%; hence labels
are reconstructed from per-candidate three-round medians rather than any
individual round.

\begin{table*}[htbp]
	\color{blue}
	\caption{Balanced LLM decode ($N{=}1$) pruning-grid sensitivity over 14
		SpMV kernels (mean over five selector seeds). Each non-overall row contains
		three complete 70/80/90\% configurations and 192 matrices. Selected time is
		the sum over that row, not end-to-end LLM latency. WISE+FFT is positive
		overall, but the block-4 activation-weighted row is negative. Bold is used
		only when the selected-time direction is consistent across all five seeds;
		within such rows it marks the higher mean accuracy and lower mean time.}
	\label{tab:llm-pruning-balanced-grid}
	\centering
	\scriptsize
	\resizebox{\textwidth}{!}{
		\begin{tabular}{llrrccccc}
			\toprule
			Granularity & Score & Configs & Matrices &
			\multicolumn{2}{c}{Tolerant Acc.} &
			\multicolumn{2}{c}{Selected Time (ms)} & Speedup \\
			\cmidrule(lr){5-6}\cmidrule(lr){7-8}
			 & & & & WISE & WISE+FFT & WISE & WISE+FFT & \\
			\midrule
			Global   & Magnitude       & 3 & 192 & 75.31\% & \textbf{86.46\%} & 0.8718 & \textbf{0.7658} & 1.1384$\times$ \\
			Global   & Activation-wtd. & 3 & 192 & 69.06\% & \textbf{89.48\%} & 0.8053 & \textbf{0.5844} & 1.3780$\times$ \\
			Block 4  & Magnitude       & 3 & 192 & 98.65\% & 98.96\% & 1.8818 & 1.8816 & 1.0001$\times$ \\
			Block 4  & Activation-wtd. & 3 & 192 & \textbf{98.96\%} & 97.50\% & \textbf{1.7460} & 1.7525 & 0.9963$\times$ \\
			Block 8  & Magnitude       & 3 & 192 & 94.58\% & \textbf{96.88\%} & 1.9324 & \textbf{1.9248} & 1.0040$\times$ \\
			Block 8  & Activation-wtd. & 3 & 192 & 93.85\% & \textbf{94.90\%} & 1.7509 & \textbf{1.7459} & 1.0028$\times$ \\
			Block 16 & Magnitude       & 3 & 192 & 94.69\% & \textbf{99.79\%} & 1.5324 & \textbf{1.5160} & 1.0108$\times$ \\
			Block 16 & Activation-wtd. & 3 & 192 & 93.85\% & \textbf{99.17\%} & 1.4934 & \textbf{1.4770} & 1.0111$\times$ \\
			Block 32 & Magnitude       & 3 & 192 & 96.46\% & \textbf{98.96\%} & 1.5004 & \textbf{1.4965} & 1.0026$\times$ \\
			Block 32 & Activation-wtd. & 3 & 192 & 97.60\% & \textbf{98.33\%} & 1.4612 & \textbf{1.4598} & 1.0009$\times$ \\
			Block 64 & Magnitude       & 3 & 192 & 99.58\% & \textbf{99.90\%} & 1.4716 & \textbf{1.4707} & 1.0006$\times$ \\
			Block 64 & Activation-wtd. & 3 & 192 & 95.62\% & \textbf{97.08\%} & 1.4451 & \textbf{1.4428} & 1.0016$\times$ \\
			\midrule
			Overall & Both & 36 & 2,304 & 92.35\% & \textbf{96.45\%} & 17.8923 & \textbf{17.5180} & \textbf{1.0214$\times$} \\
			\bottomrule
		\end{tabular}
	}
\end{table*}

Across the complete grid, WISE+FFT raises tolerant accuracy by 4.10 percentage
points and reduces aggregate selected-kernel time by 2.09\%, yielding
1.0214$\times$ speedup (1.0207$\times$--1.0220$\times$ across seeds).  The
lowest seed-specific 95\% source-layer-bootstrap lower bound is
1.0148$\times$, and deleting any one of the 36 configurations leaves every
seed above 1.0$\times$ (minimum 1.0138$\times$).  Four oracle kernels remain;
CSR is the largest class at 47.7\%, so the task is not a single-kernel
degeneracy.

The breakdown shows why a single favorable block size is insufficient.
Global masks provide the largest headroom, block 16 is consistently positive,
and blocks 8, 32, and 64 are small-positive.  In contrast, block-4 magnitude
is neutral and block-4 activation-weighted is negative at 0.9963$\times$.
The two global rows account for 87.3\% of the total time saved; after excluding
them, the pooled block-only speedup is only 1.0029$\times$.  Magnitude and
activation-weighted score-family marginals are both positive at
1.0149$\times$ and 1.0283$\times$, respectively.
All three sparsity marginals remain positive (1.0245$\times$ at 70\%,
1.0156$\times$ at 80\%, and 1.0240$\times$ at 90\%).  Held-out-model means are
also positive but heterogeneous: 1.0041$\times$ for Llama, 1.0119$\times$ for
Mistral, 1.0420$\times$ for OPT, and 1.0432$\times$ for Qwen.

\noindent\textbf{Scope clarification.}  Table~\ref{tab:llm-pruning-balanced-grid}
is a balanced exploratory sensitivity analysis introduced after the complete
task had been inspected.  It supersedes the preserved favorable-slice result as
the current exploratory analysis but is not used as confirmatory evidence.  The
fixed-protocol 2,112-matrix analysis in Section~\ref{sec:llm-application} remains the headline LLM transfer result.  Both results measure only
selected sparse projection-kernel time and exclude feature extraction,
selector inference, dense layers, attention, KV-cache management, framework
overhead, and pruning-quality effects.
}

\end{document}